\documentclass[preprint,12pt]{elsarticle}

\usepackage{epsfig}
\usepackage{amssymb}
\usepackage{amsmath}
\usepackage{xcolor}
\usepackage{listings}

\def\U{\mathbf{U}}
\def\u{\mathbf{u}}
\def\bnabla{\mathbf{\nabla}}

\newcounter{bla}

\journal{Computer Physics Communications}

\begin{document}

\begin{frontmatter}



\title{A vectorized Navier-Stokes ensemble direct numerical simulation code for plane parallel flows}

%% \title{A \LaTeX{} template for CPC Computer Programs in Physics (CPiP) articles}


\author[a]{Marios-Andreas Nikolaidis\corref{author}}

\cortext[author] {Corresponding author.\\\textit{E-mail address:} mnikolaidis@phys.uoa.gr}
\address[a]{National and Kapodistrian University of Athens, Department of Physics, \\ Panepistimiopolis, Zografos, Athens, 157 84, Greece}
%\address[b]{Second Address}

\begin{abstract}

We present a pseudo-spectal Navier-Stokes solver for plane parallel flows (Couette/Poiseuille), that has been developed on the MATLAB programming language. The code performs direct numerical simulations (DNSs) of turbulence in 3 dimensions at low Reynolds numbers ($Re_{\tau} \approx 40-180 $),
using a programming structure that 
%involves only the main mathematical operations required to solve  
mainly employs matrix multiplications to solve 
% is intuitevely connected with 
the Navier-Stokes equations.
The novelty of this DNS code lies on the vectorization of the required operations to simultaneously advance in time an ensemble of flow realizations and the  utilization of graphics processing unit (GPU) computational resources, which offers a significant reduction of running time for simulations with adequately resolved viscous scales.   

%%% Text of abstract
%A submitted program is expected to satisfy the following criteria: it must be of benefit to other physicists, or be an exemplar of good programming practice, or illustrate new or novel programming techniques which are of importance to computational physics community; it should be implemented in a language and executable on hardware that is widely available and well documented; it should meet accepted standards for scientific programming; it should be adequately documented and, where appropriate, supplied with a separate User Manual, which together with the manuscript should make clear the structure, functionality, installation, and operation of the program.
%
%Your manuscript and figure sources should be submitted through Editorial Manager (EM) by using the online submission tool at \\
%https://www.editorialmanager.com/comphy/.
%
%In addition to the manuscript you must supply: the program source code; a README file giving the names and a brief description of the files/directory structure that make up the package and clear instructions on the installation and execution of the program; sample input and output data for at least one comprehensive test run; and, where appropriate, a user manual.
%
%A compressed archive program file or files, containing these items, should be uploaded at the "Attach Files" stage of the EM submission.
%
%For files larger than 1Gb, if difficulties are encountered during upload the author should contact the Technical Editor at cpc.mendeley@gmail.com.

\end{abstract}


\end{frontmatter}

% \linenumbers

% All CPiP articles must contain the following
% PROGRAM SUMMARY.

{\bf PROGRAM SUMMARY}
%/NEW VERSION PROGRAM SUMMARY}
  %Delete as appropriate.

\begin{small}
\noindent
{\em Program Title: }                                                                                   
\\
{\em CPC Library link to program files:} (to be added by Technical Editor) \\
{\em Developer's repository link:} https://github.com/mariosn-phys/dns-ensemble-lib \\
{\em Code Ocean capsule:} (to be added by Technical Editor)\\
{\em Licensing provisions(please choose one): AGPLv3} \\
{\em Programming language: MATLAB}                                   \\
{\em Supplementary material:}                                 \\
  % Fill in if necessary, otherwise leave out.
%{\em Journal reference of previous version:}*                  \\
%  %Only required for a New Version summary, otherwise leave out.
%{\em Does the new version supersede the previous version?:}*   \\
%  %Only required for a New Version summary, otherwise leave out.
%{\em Reasons for the new version:*}\\
%  %Only required for a New Version summary, otherwise leave out.
%{\em Summary of revisions:}*\\
%  %Only required for a New Version summary, otherwise leave out.
{\em Nature of problem(approx. 50-250 words):
Turbulence is simulated in 3 dimensions with a grid resolution that captures sufficiently the viscous scales. The Navier-Stokes equations are solved in a domain with moving boundaries (the plane parallel Couette flow) or a domain where time-dependent pressure force drives a flow with constant mass flux (the Poiseuille channel flow). A multitude of such realizations is required to study
collectively problems with stochastic excitation or varying parameters.}\\
{\em Solution method(approx. 50-250 words): The domain is discretized as a Chebyshev (in the wall-normal direction), equally-spaced (in the streamwise and spanwise directions) grid where derivations are performed with pseudo-spectral matrices. We advance in time the flow field of the wall-normal velocity and vorticity utilizing a combined Crank-Nicolson/ 3rd order Runge-Kutta scheme. The next-step fields are obtained from precalculated matrix inversions. The use of matrices allows efficient evaluation of the time-stepping for a large number of distinct flows simultaneously.}\\
  %Describe the method solution here.
{\em Additional comments including restrictions and unusual features (approx. 50-250 words): The solution procedure has been formulated in matrices which is suitable for vectorization and takes advantage of parallel computations in GPUs. This enables problems where a large number of simulations is required to achieve computational times that are viable for research ,while retaining readability for educational purposes. Size of problems is restricted by GPU memory. Time integration is only possible with a fixed time-step.  }\\
  %Provide any additional comments here.
   \\

\section{Introduction}

A vast selection of Navier-Stokes solvers are available for a plethora of programming platforms. A lot of these solvers have been written with highly efficient instructions and operate in parallel interfaces (MPI), which however might not be readily accessible to new entrants in the field of fluid mechanics.
Interpreted languages such as MATLAB or Python could provide a more direct connection of the simulation code with the Navier-Stokes equations, although this ease-of-access usually comes about at the cost of performance and as such they are not used often for simulation tasks.
The specific advantages of these languages could be utilized to reduce this gap, and particularly the efficiency of MATLAB in matrix operations. 
An important development on that front has been presented with the introduction of Graphics Processing Units (GPUs), particularly these with ehnanced compute capabilities. Acceleration of scientific programs with GPU utilization can offer significant reductions in the time required to evaluate a numerical case (e.g. \cite{Jacobsen-etal-2010,Khajeh-Saeed-Perot-2013}) and even perform high Reynolds number simulations on extensive grids \citep{Vela-Martin-Encinar-2021}, although their CUDA-MPI implementations remain quite complex as well. 

% Hunt rabbits with cannons?
These DNS codes however may not be as efficient to employ in applications where large number of simulations 
are performed in small domains, such as ensemble simulations that infer a collective evolution of a mean flow in stochastically forced flow realizations \citep{Farrell-Ioannou-2012,Farrell-Ioannou-2017-bifur} or seek extreme events \citep{Ragone-Wouters-Bouchet-2018}, investigations on the properties of modified Navier-Stokes equations where various terms are controlled by a parameter (e.g \citep{Nikolaidis-alpha-2022}) and studies focused on 
obtaining multiple characteristic disturbances of a mean flow (e.g partial bases comprised by modal or optimal disturbance and Lyapunov vectors, \citep{Hogberg-2001-thesis,Nikolaidis-etal-Madrid-2018,Lozano-Duran-etal-2021}) or a base flow (e.g. \citep{Keefe-etal-1992,Inubushi-etal-2015,Nikolaidis-Ioannou-2022}).
For these types of problems a parallelized time-stepping numerical scheme of individual simulations could reduce execution times significantly,
 as shown by \citep{Jiang-2015,Jiang-Yang-2024} for typical 2-D and 3-D flow configurations.
This approach is equivalent to vectorization of the simulation ensemble members, which we can envision more intuitively in a matrix formulation.

Such features are readily supported in MATLAB, which is considered in this work since we have built a DNS code formulated with matrices, and as such it will serve  as the starting point for an ensemble DNS. 
Previous efforts to develop a MATLAB simulation code that utilizes the built-in advantages of the language
were investigated by \cite{Vuorinen-Keskinen-2016} and lead to the development of the DNSLab MATLAB suite. A finite-difference grid discretizion was used, where derivatives were defined as Kronecker matrices to skip iteration for-loops. 
Since their intention was to demonstrate that such a task can be managed without particular optimization of the simulation script,
the resulting %slower 
performance 
% who produced a DNS with reasonable performance when 
compared to openFOAM was reasonably satisfactory. 
The DNSLABIB LES Solver that employs GPUs in MATLAB was recently developed by the same Aalto CFD group \citep{Korhonen-etal-2022} as an extension of the DNSLab, which demonstrated the potential gains from such a transition, but the focus shifted to different flow configurations.

%Spectral methods
An alternative to the finite-difference (FD) method employed in the DNSLab project are the spectral method approximations where solutions to ODEs are obtained with Galerkin projections on a series of functions of increasing order \citep{Gottlieb-Orszag-1977}. 
% and their closely related pseudo-spectral \citep{Orszag,Gottlieb-Orszag,Fornberg} (or collocation) methods.
These result in highly accurate solutions at domains with simple geometries
% These are characterized by high accuracy and are considered cost-efficient, as their spectral methods counterparts (Orszag 1972).
and thus spectral method simulations in plane channel flows could be parametrized to run with less points than their counterpart FD simulations.
Their closely related pseudo-spectral methods share these spectral properties \citep{Orszag-1972}, where the equations are enforced at collocation points instead of projections on a function basis 
(\citep{Gottlieb-Orszag-1977,Lanczos-1956,Hussaini-etal-1989,Fornberg-1996})

%An attempt to reduce the gap 
In this work we present a direct numerical simulation solver of turbulent flow ensembles in a plane parralel channel, using pseudo-spectral methods. 
We aim to capitalize on the advances in computing with GPUs to achieve the necessary performance gains which can qualify this code as a realistic tool for the studies listed above than just a programming curiosity. 
Simultaneously, retaining readability and direct access to the processes of the simulation could serve 
to familiarize new entrants in the field and to translate the code to other languages, if such a need arises.

We will utilize a different solving method of the Navier-Stokes compared to DNSLab which involves a series of stored pre-calculated matrices that are stacked across the diagonal of large arrays.
%and thus significant runtime reduction. 
Although low-storage methods of RK3 time-stepping \citep{Spalart-etal-1991} are usually preferable, the present RK3 method requires the storage of only two sets of matrices as the final two steps are evaluated at the same time. The current format 
%, where the efficiency of the algorithm has been the focus of the update iterations
manages a 2.5-fold decrease in execution times of a single direct numerical simulation (DNS) on MATLAB, when the script utilizes explicitly the central processing unit (CPU). 
%, owing to the less fragmented execution of the processing operations. }
% and no 3d turbulence channel was provided.
Critically, the formulation of most operations as matrix multiplications is ideal to process them on GPUs and perform them simultaneously over many flow realizations. This allows us to construct an ensemble solver for low Reynolds number plane channel turbulence simulations that halves the time required compared to the serial solution of each flow realization.

\section{Navier-Stokes in wall-normal velocity and vorticity formulation}
\label{sec:str}

%\begin{figure}
%\includegraphics[scale=•]{}
%\caption{•}
%\end{figure}

The Navier-Stokes equations (NSE) govern the evolution of the velocity field components as a set of partial differential equations. 
The milestone of solving numerically the NSE with sufficiently resolved scales has since provided a valuable tool for the detailed analysis of turbulence \citep{Kim-etal-1987}. The NSE between parallel plates in a frame of reference with streamwise, wall-normal and spanwise coordinates $(x,y,z)$ are comprised by the system of partial differential equations for the components of the three-dimensional velocity field , the incompressibility condition and the no-slip boundary condition at the walls with coordinates $y = \pm y_w$,
\begin{eqnarray}
\partial_t \u +\u \cdot \bnabla \u + \bnabla  P -  \frac{1}{Re} \Delta \u = \mathbf{f}, \\
\bnabla \cdot \u = 0, \\
U(x,\pm y_w,z)=U_w, v(x,\pm y_w,z) = \partial_y v|_{(x,\pm y_w,z)} = 0.
\end{eqnarray}
The flow is considered periodic in the streamwise and spanwise directions, with domain size $[L_x,L_y,L_z]$ that defines the wavenumbers $k_x =  2 \pi n/L_x $ and $k_z = 2 \pi m/L_z $. % Other boundary conditions may be supplemented for each specific problem.
The non-dimensional form of the NSE is considered, with the Reynolds number $Re = h U / \nu$. The parameters $h$ (channel half-width), $U$ (characteristic velocity at the wall or centerline) and $\nu$ (kinematic viscosity) dictate the geometric similarity of the flows. 
One method to obtain numerical solutions for the NSE was described in the seminal work of \cite{Kim-etal-1987}, who derive
the equations for the Laplacian of wall-normal velocity, $\Delta v$, and the wall-normal vorticity $\eta_y$. 
This formulation utilizes the incompressibility condition to eliminate the pressure from the equations. Starting from the primitive velocity fields $(u, v, w)$, incompressibility implies that the divergence of the equations will be equal to zero and that $-\partial_y v= \partial_x u + \partial_z w$. This operation eliminates the time derivatives and the diffusive term and obtains an equation relating the advective terms with the Laplacian of pressure. The source terms of $\Delta P$ depend on the advection terms and can be readily substituted in the equation for $\Delta v$. A second dynamical variable, $\eta_y = \partial_z u - \partial_x w $, is selected since the vorticity equations resulting from the curl of NSE also eliminate the pressure. The variables $v$ and $\eta_y$ form a linear system that can be solved at each wavenumber pair to recover the respective $u$ and $w$ fields. The solutions $\u(\mathbf{x},t)$ obtained from this system are incompressible by construction.
%We can now substitute the   
%incompressiblity implies an equation between $\partial_y v$ and $u,w$ at every pair of wavenumbers. A second equation that supplements another linear condition for the wavenumbers can be found with respect to the wall-normal vorticity variable. Thus, with this pair we can reconstruct the $u,w$ fields as solutions of the linear system.

The mean profiles of $u$ and $w$, $U_{00}$ and $W_{00}$ (with $k_x=0$ and $k_z=0$), have to be explicitly calculated  and thus complete the following system
\begin{eqnarray}
\partial_t \Delta v =-(\partial_{xx}+\partial_{zz}) (\u \cdot \bnabla v)+\partial_{yx} (\u \cdot \bnabla u) + \partial_{yz} (\u \cdot \bnabla w) + \frac{1}{Re} \Delta \Delta v, \label{eq:Dv} \\
\partial_t \eta_y = -\partial_z (\u \cdot \bnabla u) + \partial_x (\u \cdot \bnabla w) + \frac{1}{Re} \Delta \eta_y, \label{eq:eta} \\
\partial_t U_{00} = \gamma(t) - \partial_y \left[ uv \right] + \frac{1}{Re} \partial_{yy} U_{00}, \label{eq:U00} \\
\partial_t W_{00} = - \partial_y \left[ wv \right] + \frac{1}{Re} \partial_{yy} W_{00}, \label{eq:W00}
\end{eqnarray}
with periodic boundary conditions at $x=0,L_x$ and $z=0,L_z$, while the following boundary conditions apply at the wall locations $(\pm y_w)$
\begin{eqnarray}
\eta_y(\pm)=0, v(\pm y_w)=0, \partial_y v(\pm y_w)= 0 \label{eq:b_ve} \\
U_{00} (\pm y_w) = \pm U_w , W_{00} (\pm y_w) = 0. 
\end{eqnarray}
The force $\gamma (t)$ is equal to 0 in Couette flow, while it imposes a constant mass flux in Poiseuille flow with $\gamma(t) = (\partial_y U_{00}|_{-y_w} - \partial_y U_{00}|_{+y_w})/Re$. 
Note that the mean profiles of $v$ and $\eta$, $V_{00}$ and $\eta_{00}$, are equal to zero at every wall-normal point.
%,  $V_{00} = 0$ and $\eta_{00} = 0$. 
\begin{lstlisting}[caption=Generation of the Inverse Laplacian for all prescribed $(k\; l)$ wavenumber pairs. \label{lst:cod_I}]
    kslov=1:NX/3+1;,msolv=1:MZ/3+1;
	
    DYF=cheb(N+1); DYF=flip(flip(DYF,1),2); I=eye(N);
    D2F=DYF*DYF; D2=D2F(2:end,2:end)
    for ii=ksolv, kp=k(ii);  
    for kk=1:length(l), lp=l(kk);
     
    alpha=kp^2+lp^2; DEL=D2-alpha*I;
    IDELv(:,:,ii,kk)=DEL\I;
    
    end, end
  
\end{lstlisting}
It is convenient for the exposition to gather the nonlinear terms of \eqref{eq:Dv}, \eqref{eq:eta} under the single variables, $F$ and $G$ respectively, since they are treated differently to the viscous terms.
%. We now  provide a detailed description for the integration using \eqref{eq:Dv},  
These equations are solved for each streamwise and spanwise wavenumber pair, e.g. for \eqref{eq:Dv_F} we write,  
\begin{equation}
\partial_t \Delta v^{(k_x,k_z)} =-F^{(k_x,k_z)} + \frac{1}{Re} \Delta \Delta v^{(k_x,k_z)}, \label{eq:Dv_F} 
\end{equation}
that describes the forced biharmonic problem for each streamwise and spanwise wavenumber pair. 
Pseudo-spectral methods are employed to perform differentiations and inversions as matrix operations in the discrete space. Substitutions result in the Laplacian $\Delta = \partial_{yy} - k^2_x - k^2_z$, where derivatives in $y$ are replaced with Chebyshev differentiation matrices ('Functions/cheb.m'). 
For the product of the two Laplacians, $\Delta \Delta = (\partial_{yy} - k^2_x - k^2_z)^2 = \partial_{yyyy} - 2 (k^2_x+k^2_z) \partial_{yy} + (k^2_x+k^2_z)^2$, we enforce both $v$ boundary conditions \eqref{eq:b_ve} at the top and bottom walls in the fourth-order derivative ('Functions/cheb4bc.m' \cite{Weideman-Reddy-2000}).
%v_{\pm w} = \partial_y v_{\pm_w} = 0$ are supplemented for the solution of the evolution equation, which are enforced in the 4-th order derivative ('Functions/cheb4bc.m' \cite{Weideman-Reddy-2000}) resulting from multiplication of the two Laplacian operators, $\Delta \Delta = (\partial_{yy} - k'^2_x - k'^2_z)^2 = \partial_{yyyy} - 2 (k'^2_x+k'^2_z) \partial_{yy} + (k'^2_x+k'^2_z)^2$. 
%With the addition of boundary conditions , we can invert $\Delta$ to obtain the $v$ equation.
We obtain $v$ from equation \eqref{eq:Dv_F} by the inverse Laplacian operator $\Delta^{-1}$ which denotes the inversion of the respective matrix representation of $\Delta$ (denoted as $DEL$ (List. \ref{lst:cod_I}))
%inverting the matrix of the Laplacian operator $\Delta$ 
with the use of the $v$ Dirichlet boundary conditions from \eqref{eq:b_ve}.

\section{The integration scheme}

We have implemented the time-stepping method for the equation system (\ref{eq:Dv}-\ref{eq:W00}) for a single flow realization with a Crank-Nicolson (CN)/ 3rd order Runge-Kutta (RK3) scheme. 
%{ The time-stepping of this dynamical system was originally performed with a Crank-Nicolson / 2nd order Adams-Basforth integration scheme, while nowadays schemes with improved numerical stability are preferred, such as the  Crank-Nicolson (CN)/ 3rd order Runge-Kutta (RK3)  scheme which is implemented in this code. } 
To advance the flow one $dt$ unit we perform 3 intermediate steps, based on the CN type of integration. 
%To advance the flow one $dt$ unit we perform 3 iterations of the CN integration, with updated advection terms. 
In the following we will describe the single realization integration scheme for equations of the form \eqref{eq:Dv_F} and investigate the benefit of facilitating their extension to a parallel execution scheme for the ensemble members, which is achieved by vectorizing the numerical integration steps.

\subsection{The CN / RK3 time-stepping}
As part of the semi-implicit Crank-Nicolson time-step, the nonlinear terms are approximated at three half-steps. To advance the solution one $dt$ unit of time the set of equations will be progressively solved for one intermediate $dt/2$ interval and one intermediate $dt$ interval before the full $dt$ interval for each iteration. A different approximation of all the nonlinear terms (last term in each of the equations (\ref{eq:CN1}-\ref{eq:CN4}))
% $F^{(k'_x,k'_z)}_{t}$ 
is employed at each intermediate time-step. For Poiseuille flow, the pressure $\gamma (t)$ enforcing constant mass flux in Eq. \ref{eq:U00} is also updated at each step.
Employing the Crank-Nicolson time-stepping, we transform (\ref{eq:Dv_F}) to a difference equation at the initial time $t$ and its' half/full time-step succession $t+h dt$,
\begin{equation}
\frac{ v^{(k_x,k_z)}_{t+hdt} - v^{(k_x,k_z)}_{t}}{hdt}  =-\Delta^{-1} F^{(k_x,k_z)}_{t} + \frac{1}{Re} \Delta^{-1} \left[ \Delta \Delta \frac{v^{(k_x,k_z)}_{t+hdt}+v^{(k_x,k_z)}_{t}}{2}\right]. 
\end{equation}
where the boundary condition of $v$ at the wall is utilized to invert $\Delta v$.
As an example,  $F^{(k_x,k_z)}_{t}$  will be evaluated at the first intermediate steps as 
$F^{0,(k_x,k_z)}_{t}$ and $h=1/2$, at the second intermediate step as $2F^{1,(k_x,k_z)}_{t}-F^{0,(k_x,k_z)}_{t}$ and $h=1$, before evaluating the final step with $(F^{0,(k_x,k_z)}_{t}+4F^{1,(k_x,k_z)}_{t}+F^{2,(k_x,k_z)}_{t})/6$ and $h=1$.
The i superscript in $F^{i,(k_x,k_z)}_{t}$ indicates the iteration of the velocity field used to calculate it, where $i=0$ denotes the actual velocity field at $t$.
Notice that although the forcing terms are updated at each intermediate step, the initial state remains always the state at $t$. 

\begin{lstlisting}[caption=x-Differentiation function of ensemble state velocity component ordered as  $(y\;x\;z\;n)$. Arrays D1x and D2x are the first and second order derivatives in x. \label{lst:cod_Dx}]

function [dFdx] = difX_Fn(F,n,Fn)

% Ensemble state dx derivatives

global D1x D2x N NX MZ

if n==1, Dif=D1x; elseif n==2, Dif=D2x; end

dFdx=permute(reshape(Dif*reshape(permute(F,[2 1 3 4]),...
[NX,(N+2)*MZ*Fn]),[NX,N+2,MZ,Fn]),[2 1 3 4]);

end

\end{lstlisting}

The $F$ term contains all nonlinear terms and derivatives in $x$,$y$ and $z$. In addition to the Chebyshev matrices for $y$ derivatives, we prescribe $x$ and $z$ derivatives with Toeplitz matrices \citep{Trefethen-2000}.
It is important to replace the for-loops that could be employed to perform the derivations with matrix operations on the appropriate dimensions of the velocity variables. For $y$  differentiations of the i-th velocity field component realizations, we reshape the four dimensional array of the velocity field into a 2D matrix by compacting the remaining dimensions into a single one, multiply by the $D(m)y$ matrix, with $m =1,2$ the derivative order, and return the differentiated field to its original shape.
For differentiations in $x$ and $z$ we include a permutation  step that transposes the differentiation coordinate to the first dimension of the matrix and restore the original order after the operation is complete (List. \ref{lst:cod_Dx}). 
After these operations the ensemble multiplication of the advection term
is calculated directly, 
since $\u \cdot \bnabla u_i$
 dictates a point by point multiplication of the $\u$ and $\bnabla u_i$ four dimensional arrays.

\subsection{Solution with precalculated inverse matrices} 
%- employment of an efficient GPU structure}

We are now tasked with allocating separately the unknown and known quantities, which result in a linear system for $v^{(k'_x,k'_z)}_{t+hdt}$,
\begin{eqnarray}
\left[ I - hdt \frac{1}{2Re} \Delta^{-1}  \Delta \Delta \right] v^{(k_x,k_z)}_{t+hdt}   = \left [ I + hdt \frac{1}{2 Re} \Delta^{-1}  \Delta \Delta \right] v^{(k_x,k_z)}_{t} \nonumber \\ 
-hdt\Delta^{-1} F^{(k_x,k_z)}_{t}, \label{eq:CN1} \\ 
\left[ I - hdt \frac{1}{2Re} \Delta  \right] \eta_{y,t+hdt}^{(k_x,k_z)}   = \left [ I + hdt \frac{1}{2 Re} \Delta \right] \eta_{y,t}^{(k_x,k_z)} \nonumber \\ 
-hdt G^{(k_x,k_z)}_{t}, \label{eq:CN2} \\
\left[ I - hdt \frac{1}{2Re} \partial_{yy}  \right] U_{00,t+hdt}   = \left [ I + hdt \frac{1}{2 Re} \partial_{yy} \right] U_{00,t} \nonumber \\ 
-(f_+(U_{+w})+f_-(U_{-w}))-hdt \gamma(t) - hdt\partial_y \left[ uv \right]_{t}, \label{eq:CN3} \\
\left[ I - hdt \frac{1}{2Re} \partial_{yy} \right] W_{00,t+hdt}   = \left [ I + hdt \frac{1}{2 Re} \partial_{yy} \right] W_{00,t} \nonumber \\ 
-hdt \partial_y \left[ wv \right]_{t}, \label{eq:CN4}
\end{eqnarray}
%These systems are linear to the variable in the left hand side and are solved directly by inverting the lhs matrix.  

These difference equations are now in the form Ax = b, with x the unknown variable at time $t+h dt$ and b the evaluated rhs of each equation. The A matrices are constant for each $h$ and thus it is significantly faster to precalculate their inverses and store them. Following the ideas of preconditioning \citep{Hussaini-etal-1989}, since there are also constant matrices in the right hand side of the difference equations, it will be economical to perform the multiplication of the inverses with the constant matrices 
and eventually calculate only the results of matrix and vector multiplications. Special care has to be applied to the $U_{00}$ equation, when the boundary conditions are non-zero (Couette). We then have to separate the first and last column of $\left[ I - hdt \frac{1}{2Re} \partial_{yy}  \right]$, which multiply the boundary condition values at the top and bottom walls respectively, and move them to the right hand side with the terms to be inverted. In the other three cases the inversion matrix is evaluated with zero boundary conditions (imposed by truncating the first and last row and column of the differentiation matrix, cf. List. \ref{lst:cod_ICv}). After clarifying the boundary conditions, we invert the A matrices in the lhs excluding the boundary points, i.e. the inversion is performed on the matrix $[I-hdt\cdots]_{(2:M_y-1),(2:M_y-1)}$. This result directly scales to $A[x_1,x_2,...,x_n]=[b_1,b_2,...,b_n]$ for n ensemble states, an equivalent relation to the one derived in the ensemble solver of \citep{Jiang-Yang-2024}.

\begin{lstlisting}[caption=Construction of the precalculated matrices after inversion of the lhs matrix in Eq.  \eqref{eq:CN2}. \label{lst:cod_ICv}]

    for ii=ksolv, kp=k(ii);
    for kk=1:length(l), lp=l(kk);
     
    alpha=kp^2+lp^2; DEL=D2-alpha*I;

    Scn_m=I-g*DEL;
    Scn_p=I+g*DEL;
        
    ICg(:,:,ii,kk)=Scn_m\I;
    ICgCg(:,:,ii,kk)=ICg(:,:,ii,kk)*Scn_p;
    
    end, end

\end{lstlisting}

%Orszag 
% This condition produces two additional vectors formed by the first and last row of the matrix A multiplied by the U velocity at the bottom and topm wall respectively 
Evaluating the system of equations on a wavenumber pair basis reduces the inversion matrices to the square $N^2_y$ of the wall-normal grid points. 
Pseudo-spectral methods apply a 2/3rds de-aliasing filter in flow fields, where the highest 1/3 of the streamwise and
spanwise harmonics are discarded \citep{Orszag-1971}.
%Each matrix solves 
The equations governing $v$ and $\eta$ are then reformulated using the compact sparse matrices with $k_x \otimes k_z \otimes R^{N^2_y }$, where the solver matrices for each ($k_x,k_z$) pair are concatenated as blocks across the diagonal (List. \ref{lst:cod_SICv}).
%{\color{yellow}, which is executed as a single matrix multiplication operation}. 
%the available speed of the GPU (or a multicore CPU). 
After reshaping $v$ and $\eta$ to single column vectors with $k_x \otimes k_z \otimes y$ Kronecker structure, we acquire their next step iteration from multiplication with these matrices (List. \ref{lst:cod_sdt}). 

\begin{lstlisting}[caption=Compilation of the sparse matrix block used in the update step of $\eta$. \label{lst:cod_SICv}]
	Nksolv=length(ksolv);
	mrange=[msolv MZ-flip(msolv(2:end))+2];Nmsolv=length(mrange);

    ICgkron=spalloc(N*Nksolv*Nmsolv,N*Nksolv*Nmsolv,...
    N*N*Nksolv*Nmsolv);       
    for ii=ksolv, ICgkronb=ICg(:,:,ii,1);
    for jk=mrange(2:end)
    ICgkronb=blkdiag(ICgkronb,ICg(:,:,ii,jk)); end
        
    ei=zeros(Nksolv,1);ei(ii)=1;
    IM=spdiags(ei,0,Nksolv,Nksolv);
    ICgkron=ICgkron+kron(IM,ICgkronb); end  


\end{lstlisting}

It is straightforward to obtain the wall-normal dependence for the Fourier coefficients of $u$, $w$ with wavenumbers $(k_x,k_z)$ from the linear system comprised by the equations used to relate $u$ and $w$ with $\partial_y v$ (incompressibility) and $\eta$ (vorticity),
%
%\begin{equation}
%\begin{pmatrix}
%
%\end{pmatrix}
%\end{equation}
\begin{equation}
\begin{pmatrix}
 -ik_x \hat{u} & -ik_z \hat{w} \\ -i k_z \hat{u} & i k_x \hat{w} \end{pmatrix} = 
\begin{pmatrix}
\partial_y \hat{v} \\ \hat{\eta}_y 
\end{pmatrix}
\end{equation}
Solutions are found in the form 
\begin{eqnarray}
\hat{u} =  \frac{-i k_z \hat{\eta}_y + i k_x \partial_y \hat{v}}{k^2_x+k^2_z}  \\
\hat{w} = \frac{i k_x \hat{\eta}_y + i k_z \partial_y \hat{v}}{k^2_x+k^2_z}
\end{eqnarray}
which are also stored with order $k_x \otimes k_z \otimes y$ in the sparse matrices $\mathbf{kkm}$ and $\mathbf{llm}$, with the  elements $\frac{i k_x}{k^2_x+k^2_z}$ and $\frac{i k_z}{k^2_x+k^2_z}$ filling the diagonal. 

Lastly, the mean profiles $U_{00}$ and $W_{00}$ are evaluated separately. With $U_{00}(y)$ and $W_{00}(y)$ we determine the $u$ and $w$ components of the velocity fields and thus acquire the complete solution for the next intermediate iteration or time-step.

%We can now write these steps in discrete form according to the CN integration, which evaluates the viscous term as the mean value at the initial and following time-step.
 
\begin{lstlisting}[caption=Evaluation of the updated Fourier coefficients on the first RK3 step for all de-aliased wavenumbers of $\eta$(g1v). \label{lst:cod_sdt}]

ghat=fft2_cuben(g0,Fn);
g1v=reshape(permute(ghat(2:end-1,ksolv,mrange,:),...
[1 3 2 4]),[N*(Nksolv)*(Nmsolv),Fn]);
ag1hat=fft2_cuben(advg1,Fn)*(-dt)*h;
ag1v=reshape(permute(ag1hat(2:end-1,ksolv,mrange,:),...
[1 3 2 4]),[N*(Nksolv)*(Nmsolv),Fn]);

g1v=ICgkron1*ag1v+ICggkron1*g1v;

\end{lstlisting} 
 
\section{Evaluation of the numerical code}

\begin{figure}
\centering
\includegraphics[width= 1\textwidth]{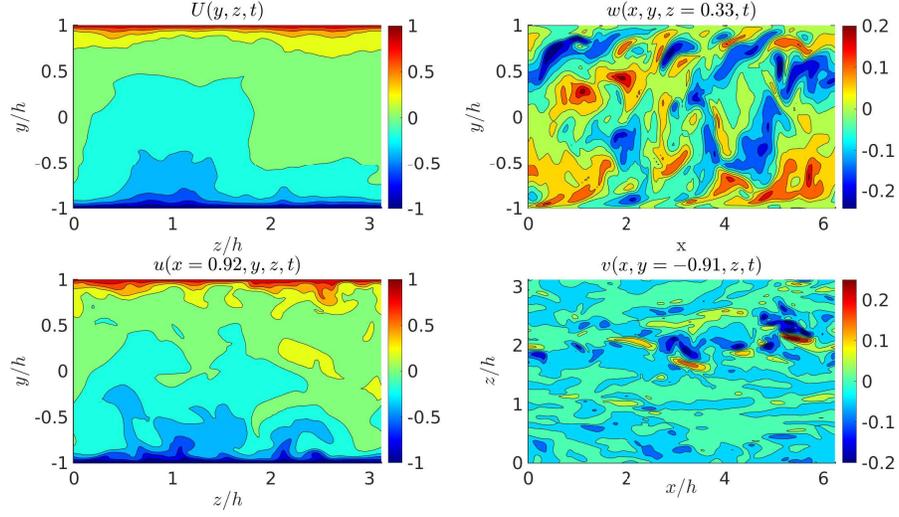}
\caption{ Instantaneous snapshots from the Couette test case C3 (cf. table \ref{table:stats}) of the mean streamwise velocity $U(y,z,t)$ and velocity fields $(u,v,w)$ at $y-z$, $y-x$ and $z-x$ cross-sections on selected planes. \label{fig:snap_C3}}
%The rms profiles from C3000 (red squares) and \citep{Pirozzoli-etal-2014} (blue lines) of the streamwise ($u'/u_{\tau}$), wall-normal ($v'/u_{\tau}$), spanwise ($w'/u_{\tau}$) velocity fluctuations from the mean profile and the tangential Reynolds stresses $u'v'/u^2_{\tau}$ scaled in wall-units.}
\end{figure}

\begin{figure}
\centering
\includegraphics[width= 1\textwidth]{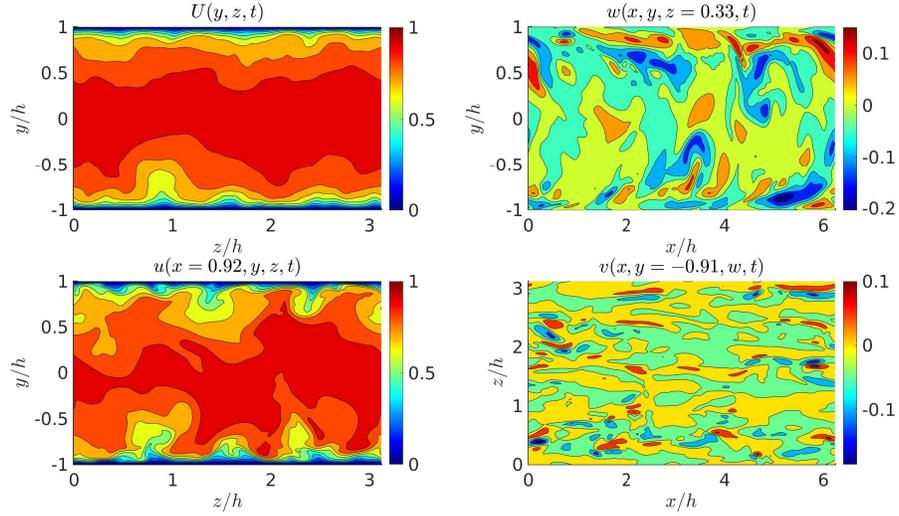}
\caption{ Same as Fig. \ref{fig:snap_C3} for the Poiseuille test case P3 (cf. table \ref{table:stats}). \label{fig:snap_P3}}
%The rms profiles from C3000 (red squares) and \citep{Pirozzoli-etal-2014} (blue lines) of the streamwise ($u'/u_{\tau}$), wall-normal ($v'/u_{\tau}$), spanwise ($w'/u_{\tau}$) velocity fluctuations from the mean profile and the tangential Reynolds stresses $u'v'/u^2_{\tau}$ scaled in wall-units.}
\end{figure}

\begin{center}
\begin{table}
%\caption{}
\caption{\label{table:stats} Simulation statistics of Couette (C2,C3,C4) and Poiseuille (P2,P3) test cases. Dimensions of the channels in wall units are indicated by $[L_x^+,L_y^+,L_z^+]$. 
$R_{\tau} = h u_{\tau} / \nu$ is the friction velocity Reynolds number according to $u_{\tau}$.
The average spacing in $x$ and $z$ for the resolutions of simulations
% with designations 2 and 3 
(cf. table \ref{table:geometry3}) are denoted as $\Delta x ^+,~\Delta z^+$. Wall-normal spacing near the wall and at the centerline is denoted as $\Delta y ^+$. The averaging time, $T u_t / h$, is also reported. }
\centering\vspace{.8em}
\begin{tabular}{@{}*{7}{c}}
\break
 Abbreviation  &$[L^+_x\;\;,\; L^+_y,\;\; L^+_z]$& $R_{\tau}$ & $Tu_{\tau}/h$ & $\Delta x^+$ & $\Delta z^+$ & $\Delta_y^+$ \\
 C2   & $[1073.8\;,341.8,\;536.9]$ & $170.9$ & $267.7$ &   $15.1$ & $7.6$ & $0.16 - 7.5$ \\
 C3  & $[1076.3\;,342.6,\;538.2]$ & $171.3$ & $262.6$ & $11.3$ & $5.7$ & $0.13 - 6.6$ \\
 C4   & $[2145.1\;,341.4,\;1072.5]$ & $170.7$ & $273.1$ &   $20.8$ & $10.4$ & $0.13 - 6.5$ \\
 P2   & $[1157.4\;,368.4,\;578.7]$ & $184.2$ & $221.0$ &   $16.3$ & $8.2$ & $0.18 - 8.0$ \\
 P3  & $[1150.5\;,366.2,\;575.2]$ & $183.1$ & $276.0$ & $12.1$ & $6.1$  & $0.13 - 7.0$ \\
\end{tabular}
%\label{table:geometry}
\end{table}
\end{center}

As validation for our code we collect one point statistics of the solution velocity fields.  These are the time-averaged streamwise- and spanwise-mean velocity profile $U(y)$, the rms profile of the velocity fluctuations from the mean profile $(u',v',w')$ and the tangential Reynolds stress $u'v'$. For the statistics we will run extensive DNSs using the GPU option of the code. 
Transition from CPU to GPU computation is implemented by the simple transfer of the appropriate variables to the GPU with the MATLAB command gpuArray.
The well-resolved cases $(C3,P3)$ require approximately two days to complete over 250 $T u_{\tau}/h$ units of time. 
Instances of the mean streamwise velocity and cross-sections of the velocity fields $(u,v,w)$ maintained under these conditions are shown in Figs. \ref{fig:snap_C3} and \ref{fig:snap_P3}.
We can compare the agreement of our statistics with the statistics of \citep{Pirozzoli-etal-2014} (for plane Couette flow)  and \citep{DelAlamo-Jimenez-2003} (for plane Poiseuille flow). This comparison is shown for the cases of table \ref{table:stats} in Figs. \ref{fig:umeanc} and \ref{fig:statc}. 

It is evident that both the mean profile statistics and the velocity fluctuation rmss agree considerably with the benchmark $R_{\tau}=171$ case presented by \citep{ Pirozzoli-etal-2014} given that their case was calculated in a bigger domain. We can verify that statistics away from the wall converge to those of \citep{ Pirozzoli-etal-2014} by running the low resolution C4 in a box with periodic dimensions $[L_x, L_z]=[4 \pi, 2 \pi]$.    
%Quality of the solutions / Matching Statistics
Similar agreement was found when we compared the Poiseuille mean profile and fluctuation statistics with the benchmark of \citep{DelAlamo-Jimenez-2003}.
Instances of the streamwise-mean $U$ velocity field show the characteristic streaks rising from the walls in both flow configurations (Figs. \ref{fig:snap_C3}a and \ref{fig:snap_P3}a)

\begin{figure}[h]
\centering
\includegraphics[width= 0.9\textwidth]{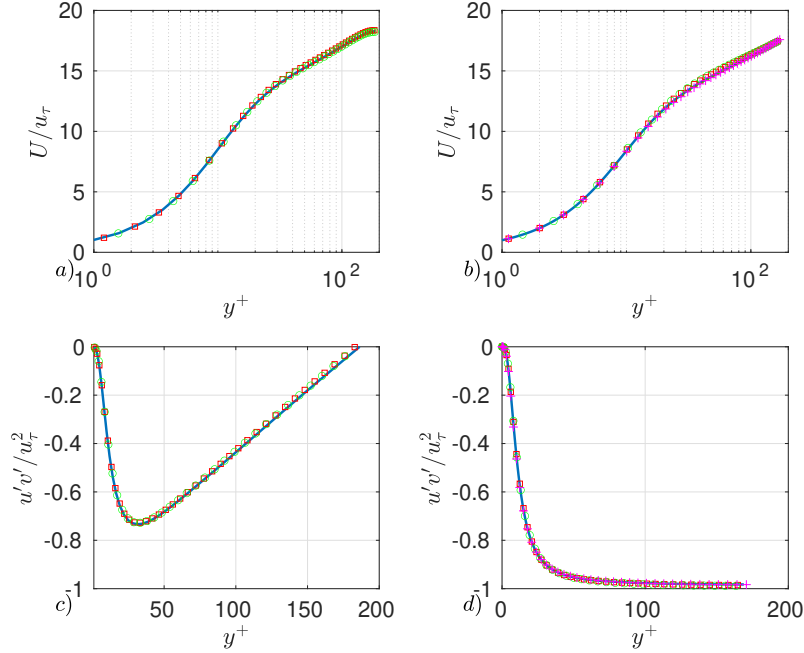}
\caption{ (a,c) Comparison of P2 (red squares) and P3 (green circles) with the statistics of the mean velocity profile $U$ (a) and the tangential shear stress (c) \citep{DelAlamo-Jimenez-2003}(solid blue lines). (b,d)  
Comparison of C2 (red squares), C3 (green circles) and C4 (purple crosses) with the statistics of the mean velocity profile $U$ (b) and the tangential shear stress (d)
of \citep{Pirozzoli-etal-2014} (solid blue lines). The results from our code show good agreement in both cases.
%Mean velocity profile $U$ (a,b) in wall-units and tangential shear stress $u'v'$ (c,d) of velocity fluctuations from the mean profile.
%. (a) Comparison of C2 (red squares) and C3 (green circles) with the statistics of \citep{Pirozzoli-etal-2014} (solid blue line. 
\label{fig:umeanc}}
\end{figure}

\begin{figure}[h]
\centering
\includegraphics[width= 0.9\textwidth]{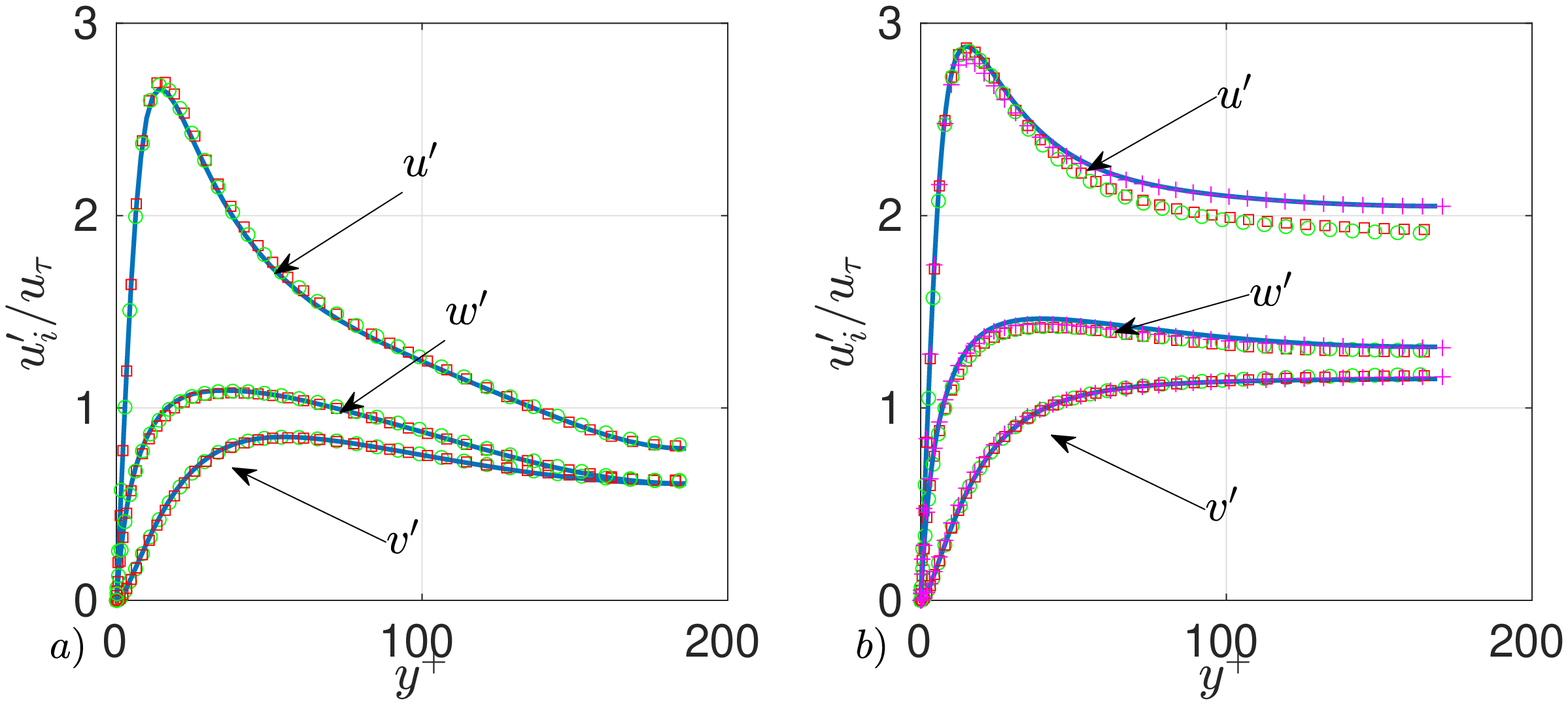}
\caption{The rms profiles from a) \citep{DelAlamo-Jimenez-2003} and b) \citep{Pirozzoli-etal-2014} ( solid blue lines) of the streamwise ($u'/u_{\tau}$), wall-normal ($v'/u_{\tau}$), spanwise ($w'/u_{\tau}$) velocity fluctuations from the mean profile $U$.These statistics agree reasonably well near the wall, whereas a deviation is noted in b) for the $u',w'$ profiles since the domain of \citep{Pirozzoli-etal-2014} is considerably larger.  This deviation further away from the wall is indeed seen to reduce for the coarse C4 case in a larger domain.
%and the tangential Reynolds stresses $u'v'/u^2_{\tau}$ 
%scaled in wall-units.
\label{fig:statc}}
\end{figure}

%\begin{figure}
%\includegraphics[scale=•]{}
%\caption{•}
%\end{figure}

\subsection{Ensemble integrations}

% Ensemble Solver

In this section we evaluate the advantages in running time for the case C1 (see table \ref{table:grid1} for parameters) with ensemble simulations comprised by 5 to 25 ensemble members. The ensemble version of C1 is run at a lower Reynolds number, $Re=1500$, to ensure a well-resolved grid for the ensemble states. The duration of these simulations is compared to a serial ensemble with the same number of modes and a parallel execution of single simulation instances. We sample a period of approximately 600 time-steps for the ensemble runs, awaiting for the performance decrease due to thermal throttling to occur before averaging.

We examine the serial ensemble first, where the ensemble solver cycles sequentially over each flow realization every time-step.  
This method obtains separately the new velocity fields at the cost of repeating N times the required numerical operations, so the only benefit in this case is lower memory usage.
As expected, the serial ensemble time-step duration is practically equal to the single simulation time-step multiplied by each respective N. This is supported in Fig. \ref{fig:Tdt_{ens}}, where the average time-step duration, $\bar{T}_{dt}$, for this case increases linearly with N. 

Another approach introducing parallelism may be pursued, whereby the simulations are launched simultaneously as separate instances. Parallel execution of simulation realizations shows an initial decrease relative to the serial case, however this simultaneous execution is limited to 5 ensemble members for our set-up since the memory load of individual simulations becomes prohibitive fast.  

%{\color{red} It is possible to integrate an N-member simulation ensemble directly by considering a sequential evaluation of the individual evolutions at every time-step (e.g \citep{Farrell-Ioannou-2017-bifur}).}

We compare these test cases with the vectorized ensemble implementation of the DNS . It is evident from the initial N=5 case that the vectorized simulations significantly reduce the execution time and the memory usage compared to the parallel runs. For a direct comparison, the average time-step time of these runs is equal to $0.3221 \pm 0.0040$ (serial-ensemble), $0.2881 \pm 0.0034$ (parallel-single) and $0.1475 \pm 0.0028$ (vector-ensemble).

%Parallel N5 $0.2881 \pm 0.0034$
%Vector N5 $0.1475 \pm 0.0028$ 
%Serial N5 $0.3221 \pm 0.0040$
 The efficiency of the solver is evident when collecting statistics of the flow for one of the single realization cases shown in the beginning of the section. An ensemble simulation of C2 with four members (optimal ensemble for the GPU memory) collects statistics in nearly half the time ($\bar{T}_{dt} = 0.347 \pm 0.007 $) required from the single realization ($4 \bar{T}_{dt}=0.524$, documented in \ref{App:time}) to gather them over the same time interval. We can initialize this simulation with four copies of a single flow state subjected to stochastic forcing over a short period of ten eddy turnovers. 
%Instances of the stochastic forcing are generated by a linear combination  of the diffusion operator eigenvectors, scaled to produce the same diffusion rate.

%Afterwards, the forcing is switched off and we advance each state for an additional 100 $T u_{\tau}/h$ units.   

\begin{figure}[h]
\centering
\includegraphics[width= 0.75\textwidth]{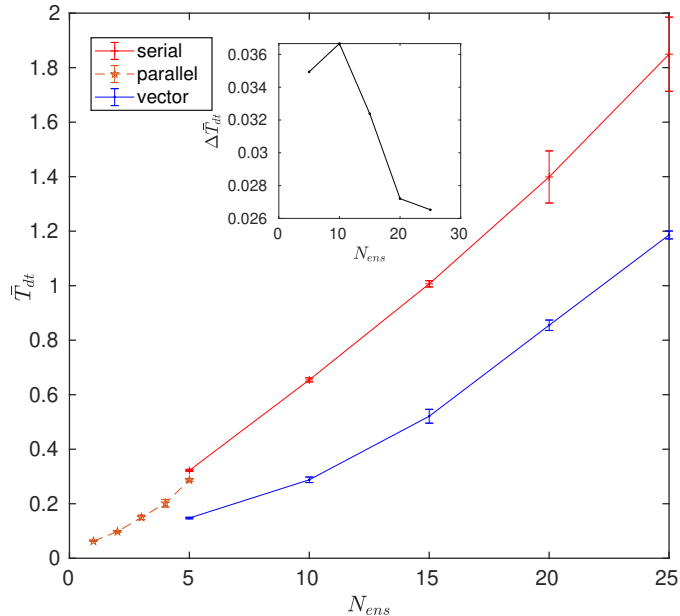}
\caption{Average execution time $\bar{T}_{dt}$ required to advance the ensemble C1 flows one time-step $dt$ as a function of ensemble size. The difference of serial and vectorized $\bar{T}_{dt}$ divided by $N_{ens}$ in the inset shows that $N=10$ is the most efficient ensemble size for the present case and hardware configuration.
\label{fig:Tdt_{ens}}}
\end{figure}

%Ensemble integrations 

In addition to statistics, we are also able to study more efficiently the dynamics of    statistical states. 
The onset of transition to turbulence is characterized by the formation of elongated streamiwse rolls and streaks. These modify the laminar profile and are responsible for the deviation from spanwise homogeneity.
Statistical state dynamics provide a theoretical explanation for the observed organization of rolls and streaks in the streamwise mean flow. In this framework, rolls and streaks emerge as an instability that breaks the symmetry of the homogeneous state when the forcing amplitude exceeds a critical threshold.
The S3T second order cumulant closure of the Navier-Stokes \citep{Farrell-Ioannou-2012} predicts the onset of this instability and its equilibration in an ensemble with infinite members.

Here, we perform an approximate S3T calculation utilizing an ensemble simulation (e.g \citep{Farrell-Ioannou-2017-bifur}) with $N_{ens}=5$ . The flows are simulated in a domain $[L_x , L_y , L_z]/h=[2 \pi, 2 , 1.2 \pi]$ with C1 resolution at $Re = 1000$. Each ensemble member shares a common streamwise mean flow, which is updated at every time-step by the perturbation Reynolds stresses averaged over the ensemble members.      
We generate the stochactic forcing realizations by mixing eigenvectors of the diffusion operator acting on a $(\Delta v, \eta)$ state, scaled to produce equal diffusion rate, with streamwise wavenumber $k_x/a=1$.
%from structures with equal dissipation  on the streamwise wavenumber $k_x/a=1$.
The energy density level of the forcing maintains a perturbation energy density equal to a $10^-3$ of the laminar energy density, which is equivalent to a velocity rms $\approx 0.025U_w$.
 The amplitude of the roll and streak in this ensemble simulation is found to grow exponentially over time with growth rate $\sigma = 0.0016$ and reach a quasi-equilibrium which is characterized by the roll and streak shown in Fig. \ref{fig:s3t_ens}(a). This approach is shown to estimate correctly the growth rate of rolls and streaks in single realization simulations (Fig. \ref{fig:s3t_ens}(b)).

\begin{figure}
\centering
\includegraphics[width= 0.95\textwidth]{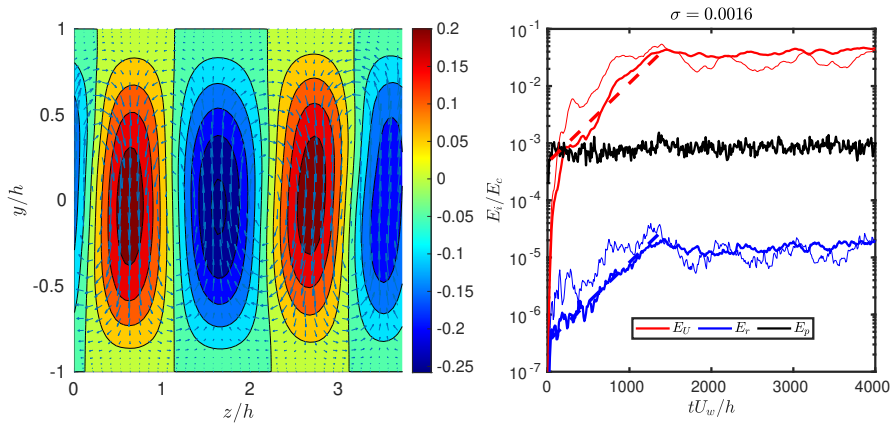}
\caption{(a) A streamwise-mean flow snapshot depicting the quasi-equilibrated S3T mode in a $N_{ens}=5$ ensemble DNS with stochastic forcing applied during the simulation. The $L_z$ size of the domain was increased to $1.2 \pi$ to accomodate a two pair S3T mode. Contour levels are plotted at 0.05 intervals and the maximum absolute values of cross-flow velocites are $max(v,w) = (0.0052,0.0032)$. (b) Energy density fractions of streak(red), roll(blue) and perturbations with $E_c$. The time-series are denoted with thick lines for $N_{ens}=5$ and with thin lines for single realization. Dashed lines indicate that the instability in $N_{ens}=5$ grows at an exponential rate $\sigma = 0.0016$.
\label{fig:s3t_ens}}
\end{figure}

%Power method diagnostics

\subsubsection{Diagnostic utilities that benefit from vectorization}

Another application where ensembling reduces computational time is in dynamical system diagnostics and control applications which require the tracking of multiple linearized states simultaneously with the numerical solution. These may tackle the acquisition of a set of Lyapunov vectors, modal distrubances or optimals about a base flow (e.g \citep{Hogberg-2001-thesis,Nikolaidis-etal-Madrid-2018,Lozano-Duran-etal-2021}. An advantage of this type of stability analysis is the  inclusion of the roll component in the mean which is usually neglected in eigenanalysis, thus the recovered modal disturbances are more representative of the actual flow. In return we can obtain only partially the eigenvector basis, which suffices for applications that aim to capture periodically all unstable eigenvectors of the mean flow or optimal disturbances above a growth threshold. 
Depending on the diagnostic, we solve one time-step concurrently with the flow integration (Lyapunov vectors e.g \citep{Nikolaidis-Ioannou-2022}) or solve both forward and adjoint equations for a fixed number of iterations (modal-adjoint vectors e.g. \citep{Lozano-Duran-etal-2021}) or for a given time window (optimal growth disturbances) that produces adequate convergence for the specific application.
This function could also be modified to track time-dependent optimals or covariant Lyapunov vectors with defined $k_x$, where we substitute the constant mean flow at the beginning with the mean flow time-series for the given interval.

\begin{figure}[h]
\centering
\includegraphics[width= 0.95\textwidth]{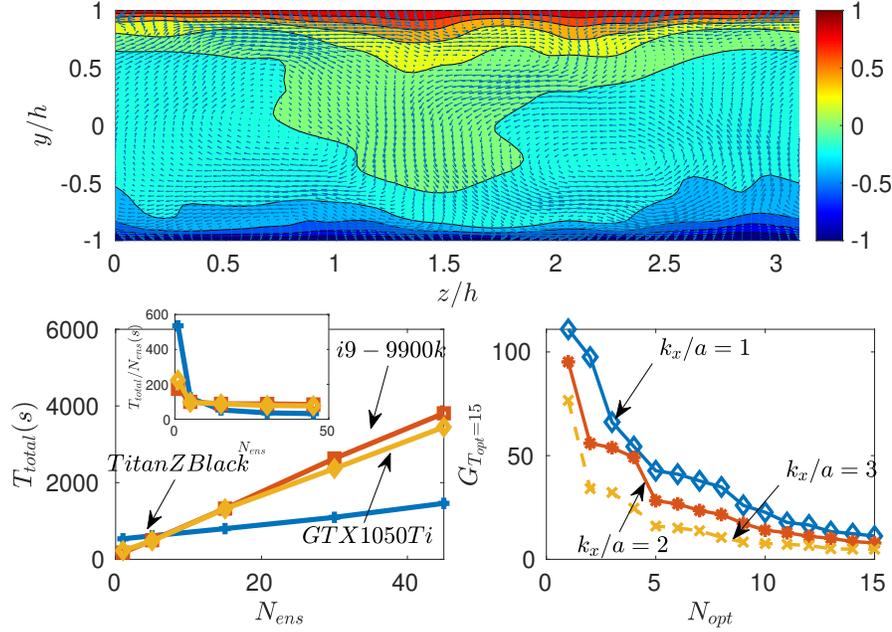}
\caption{ (a) The selected mean flow from C1 where optimization is performed. This mean flow includes wall-normal and spanwise components V,W (arrows). (b) Total execution time $T_{total}$ for an ensemble calculation of $N_{ens}$ optimal disturbances. Titan Z (blue), CPU(red),  1050 Ti (yellow).
The difference of $T_{total}$ divided by $N_{ens}$ in the inset shows that the Titan Z becomes significantly more efficient as the problem size increases compared to the other two options. 
(c) The first 15 optimal disturbances of the streamwise mean flow (a) for wavenumbers $k_x/a =[1,2,3]$ (blue, red, yellow lines) obtained in simulation time $T_{total}= 1460 $ (b). 
\label{fig:Opt_ens}}
\end{figure}

We will consider an example of finding disturbances that optimize energy growth at a target time $T_{opt}=15$ on the streamwise-mean base flow shown in Fig. \ref{fig:Opt_ens}(a)(C1 case) with forward and adjoint equations \citep{Luchini-2000}. In this particular case we may specify the streamwise wavenumber of each disturbance. 
%A particular case benefitting from vectorization are disturbances of the single harmonic form $e^{i k_x x}$ that correspond to the streamwise-averaged mean flow. 
Since these perturbations do not have a mean flow, we only solve Eqs. (\ref{eq:CN1}-\ref{eq:CN2}) for the specific wavenumber of each disturbance  to update them.                                                        
Therefore the preconditioned update matrix is compiled from the single $k_x$ precalculated matrices that correspond to each disturbance vector.The matrices are also valid for the adjoint equations.
The implementation of this solver to evaluate optimal disturbances is comprised by an iteration loop of forward and adjoint integrations over the optimization target time and an orthonormalization step. Its efficiency in calculating a large number of optimals is greatly increased by vectorization, as shown in Fig. \ref{fig:Opt_ens}(b) by the comparison between a single vector and a partial basis of vectors increasing up to 45. The inset plot in \ref{fig:Opt_ens}(b) also reveals that for single vectors the CPU is faster than both GPUs, however the trend is reversed as we increase the size of the optimal basis.    

%Optimal Disturbances spectra.
%Modal-Adjoint Stability. 
%deduce - converge - retrieve
  An observation that significantly accelerates execution of the diagnostics code is that it remains stable for larger time-steps than the original DNS. Hence, we use a $\Delta t = t_f dt$ that is multiplied by a factor $t_{f}=5$ compared to the DNS $dt$. In this way the number of steps required to evaluate one optimal loop is reduced by the same factor.For the present case, most of the top-modes and their growth rates remain fairly unchanged for $t_{f}$ ranging from 1 to 10, with exceptions attributed to the random initialization seeds and optimals with comparable growth rates (cf. Fig. \ref{fig:Opt_ens}(c)) that may result in mixing.
%May cause adjacent modes harder to distinguish  
   Further increases lead to noticeable divergence of the modes and eventually to breakdown of the higher wavenumber harmonics. Careful selection of this acceleration factor based on the stability of the higher modes for each wavenumber will retain the desired accuracy at affordable computation times.

%on the top modes of each wavenumber to ensure stability will retain the desired accuracy at affordable computation times. {\color{red} Test inner products with h=1 and h=10,20, //5 and 10 reasonably close}}
 
%  as such this optimization would probably halve      

\subsection{Ensemble timing benchmarks}

\begin{center}
\begin{table}
%\caption{}
\caption{\label{table:enTZ} Typical duration of distinct types of numerical operations (fft2 transforms, sparse matrix multiplications (smm), elementwise multiplications (em) and full matrix multiplications (fmm))that scale with the ensemble size $N_{ens}$.
The total GPU execution time on Titan Z Black for the number of calls in the parantheses, in a simulation with 800 time-steps.}
\centering\vspace{.8em}
\begin{tabular}{@{}*{6}{c}}
\break
 $N_{ens}$  &$fft2(9600)$& $smm(1600)$ & $em(2400)$ & $fmm(2400)$  \\
 1   &$3.697$& $4.13$ & $0.73$ & $2.072$  \\
 3   &$7.383$& $4.47$ & $1.41$ & $3.886$ \\
 10  &$24.73$& $9.65$ & $4.42$ & $12.341$ \\
 25  &$63.47$& $20.48$ & $10.45$ & $28.873$  \\
\end{tabular}
%\label{table:geometry}
\end{table}
\end{center}

We will now focus on the ensemble simulations and time some specific operations on the Titan Z Black and the GTX 1050 Ti GPUs to detect the source of reduction and also potential bottlenecks. The greatest reduction is seen (table \ref{table:enTZ}) in sparse matrix multiplication, whereby the 25 member ensemble is only 5 times slower than the single DNS. Notable gains are observed in the other 3 operations, with elementwise and full matrix operation reducing to $60 \%$ and fft2 to $70 \%$ of their respective serial DNS implementations.
The smaller gains for fft2 suggest that the function is significantly optimized even for single members, since the migration from CPU to GPU already reduced the duration of this operation to its 1/6-th (\ref{App:time}).
%suggest that the function is  optimized to use all threads even when a single member is considered, which is reasonable given the .}
For these operations, the main improvement is seen when switching from $N=1$ to $N=3$ and then scales linearly with ensemble size. 
The increasing cost of differentiations is associated with permutations, 
given that solely matrix multiplications such as the $y$ derivative are only marginally slowed by increasing N from 1 to 3.
This improvent applies to $N=25$ as well, though a slow-down element is introduced in the form of increased time required for memory deallocation while exiting functions.  
 Performing the benchmarks on a GTX 1050 Ti does not reveal the same benefits (table \ref{table:enTi}), which are exclusive to the sparse matrix multiplications.

%%%% 1, 3, 10, 25
%
%fft2 ~1.2 - 1
%spar ~ 2 - 10
%elem ~ 6 - 10
%diff ~ 6 - 10

\begin{center}
\begin{table}
%\caption{}
\caption{\label{table:enTi} Typical duration of distinct types of numerical operations (same as table \ref{table:enTZ}) that scale with the ensemble size $N_{ens}$ . The total GPU execution time on GTX 1050 Ti for the number of calls in the parantheses, in a simulation with 800 time-steps.}
\centering\vspace{.8em}
\begin{tabular}{@{}*{6}{c}}
\break
 $N_{ens}$  &$fft2(9600)$& $smm(1600)$ & $emm(2400)$ & $fmm(2400)$  \\
 1   &$8.958$& $9.974$ & $1.16$ & $7.197$  \\
 3   &$23.258$& $15.272$ & $3.078$ & $20.627$ \\
   10  &$78.35$& $34.566$ & $10.337$ & $70.437$ \\
\end{tabular}
%\label{table:geometry}
\end{table}
\end{center}

The Titan Z Black hardware used for these simulations would be considered enthusiast grade at its' release time and theoretically achieves a peak double precision GLOPs performance of $\approx 1050$ in multiplication of square matrices with size $2048^2$. The GTX 1050 Ti is more of an entry level GPU, with $\approx 70$ available GFLOPs.
Their comparison shows only an improvement of $~2.5 - 3 \times$ in performance instead of the expected $15 \times$. Given that we perform a variety of operations that are not sufficiently large or do not feature the optimal square matrix shape, it may be consistent that we cannot obtain the full benefit of the Titan Z Black. Thus, it could prove beneficial to investigate more compact approaches for differentiations.
These gains however will be applicable only to full matrix multiplications and given that they are responsible for about a 3rd of the time spent we estimate that the reduction would be a "modest" $20\%$ of the total time. The optimization diagnostics indicate that
newer hardware is reasonably faster in serial calculations, thus we expect that substantial improvement can be sought in more recent computational GPUs, such as the P100 or V100. These are also 
 expected to run ensembles with more grid points and more members.
% to significantly improve on the current 

% another improvement can be sought in utilizing newer hardware, such as the P100 or V100, that are also expected to run ensembles with more grid points and more members. } 

\section{Conclusions}

We have presented a Navier-Stokes solver mainly formulated with matrix operations, which utilizes the built-in MATLAB vectorization.
This code reduces the duration of time-steps in 
similarly resolved simulations with these performed by the DNSLab by $\approx 2.5\times$ when both codes use exclusively the CPU, which increases up to $\approx 9\times$ when we utilize the GPU acceleration available in our code.
DNSs spanning ~250 eddy turn-over times were completed in a time-frame between tens of hours to a couple of days given the hardware at our disposal and the resolution of the computational grid.
We also demonstrated the accuracy of simulation statistics for the two main channel options available in the code, the plane Couette and plane Poiseuille flows, at a friction Reynolds number of $\approx 170-180$.

In addition to the base DNS results, we provided details of a vectorized  ensemble version of this code, which achieves another halving of execution time when multiple simulations of the same domain are required. 
Another benefit of the ensemble solver is the lesser memory requirements per simulation realization.
This implementation facilitates efficient simulation of ensemble states and may be used to accelerate simulations in studies of stochastic excitation and transition (e.g. \citep{Farrell-Ioannou-2017-bifur} where an earlier, non-vectorized version of this code was used for the DNS and RNL ensembles), as well as extreme event identification \citep{Ragone-Wouters-Bouchet-2018} . Such vectorized ensemble  implementations have been adapted to accelerate power method  diagnostics that study dynamical system properties of low Reynolds number turbulence. 
The code can be utilized for ensembe simulations of the NSE in its current form, but may also be ported to more efficient languages given its not overly convoluted formulation.

\begin{flushleft}
Acknowledgements
\end{flushleft}

M.-A. Nikolaidis wishes to thank Petros J. Ioannou, Brian F. Farrell and Navid C. Constantinou for fruitful discussions during the development of this code and opportunities to use it.
% and for providing the GPU where this code has been developed. 
%Additional thanks are extended to J. Jim\'{e}nez for sharing the MPI DNS code developed by the Fluid Dynamics group at UPM and A. Lozano-Dur\'{a}n for assistance with its setup and operation in the First Multiflow Summer School (UPM, 2013).
Additional thanks are extended to J. Jim\'{e}nez and A. Lozano-Dur\'{a}n for introducing the author to the MPI DNS code developed by the Fluid Dynamics group at UPM. Finally, 
% during the First Multiflow Summer School (UPM, 2013).
comments of the anonymous reviewer for the previous version of this manuscript are gratefully acknowledged.

\begin{flushleft}
Declaration of Competing Interest
\end{flushleft}

The author declares that he has no known competing financial interests or personal relationships that could have appeared to influence the work reported in this paper.

%\begin{thebibliography}{0}
%\bibitem{1}Reference 1         % This list should only contain those items referenced in the                 
%\bibitem{2}Reference 2         % Program Summary section.   
%\bibitem{3}Reference 3         % Type references in text as [1], [2], etc.
%                               % This list is different from the bibliography at the end of 
%                               % the Long Write-Up.
%\end{thebibliography}
%* Items marked with an asterisk are only required for new versions
%of programs previously published in the CPC Program Library.\\
\end{small}

%\section{}
%\label{}

\appendix

%\section{Base DNS Timing Results}
%\subsection{Timing of operations on CPU, GPU and vectorized versions of the code}

\section{Timing comparisons with the MATLAB DNSLab solver \label{App:time}}

\begin{figure}[h]
\centering
\includegraphics[width= 0.75\textwidth]{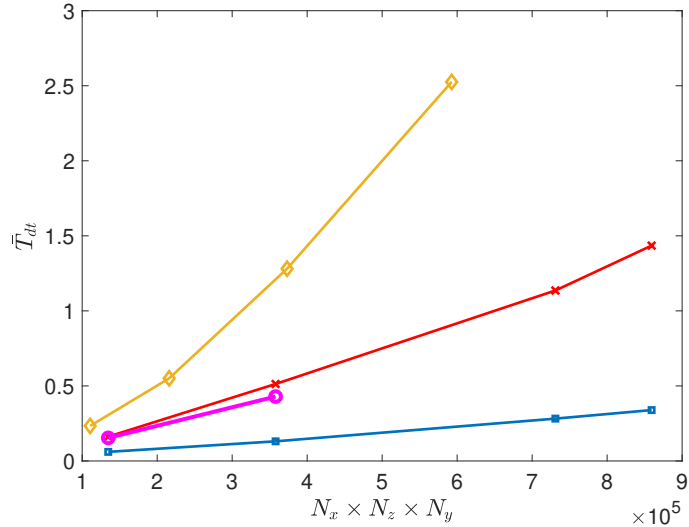}
\caption{Average execution time $\bar{T}_{dt}$ required to advance the flow one time-step $dt$ as a function of resolved grid size.
%In both the CPU (red line) and GPU (blue line) $\bar{T}_{dt}$ increases approximately as a linear function of the number of resolved grid points. 
$\bar{T}_{dt}$ increases approximately as a linear function of the number of resolved grid points for the GPU cases (blue line), while the CPU (red line) moves away from this trend at the final point.  
We compare these with the $\bar{T}_{dt}$ from a series of simulations with the DNSLab suite \citep{Vuorinen-Keskinen-2016} (mustard diamond) run at the same CPU. Finally, the purple circles are the C1-C2 cases run at a low-end GTX 1050 Ti GPU, which is still marginally faster than the respective cases run on the CPU. \label{fig:Tdt}}
\end{figure}

The base study for comparison will be the DNSlab program suite developed by \cite{Vuorinen-Keskinen-2016} for MATLAB and specifically their 3D Navier-Stokes solver. There a plane Poiseuille channel was simulated in a $[L_x \times L_y \times L_z] = [2 \pi \times 2 \times \pi]$ finite difference grid at friction Reynolds number $180$. A relatively coarse grid of $72^3$ points was employed (without including the 2 points at the walls), which demonstated the potential of MATLAB to run simulations of turbulence. 

%{\color{red} Development of our Navier-Stokes solver was also started on a finite difference grid, which was eventually revised to the current version with a Chebyshev grid that enhances resolution near the walls and a programming structure that is more suitable for GPU calculation.} 

We will initially compare the case presented by \cite{Vuorinen-Keskinen-2016}(VK3) with the C2 case computed by the present DNS code. Since we employ the 2/3rds rule, we use an increased number of points in $x$ and $z$, which after dealiasing result in a marginally smaller but comparable number of grid points (373.2k for VK3, 357.9k for C2) . 
Crucially though, for the spectral methods one half of the harmonic pairs are complex conjugate to the other half and therefore we only need to %solve one 
%we are to 
tackle matrices with half the number of elements in each dimension, i.e with size one quarter of the VK3 case.
The matrices in our code however are significantly denser due to the Chebyshev differentiation, which is reflected in the greater memory space occupied by the matrices we utilize. We expect the memory requirements to scale as $9\cdot 16[(N_x/2)\cdot N_z\cdot N_y^2]$ bytes, since 9 sparse matrices with this structure are loaded to the GPU memory, each element occupies 16 bytes due to itself and its indices, and only the sub-matrices with $N^2_y$ elements are dense. 

\begin{center}
\begin{table}
%\caption{}
\caption{\label{table:geometry3}Simulation parameters of C1-C4 and P1-P3.  
The channel  size in the streamwise, wall-normal and spanwise  directions is $[L_x,L_y,L_z]/h=[2\pi,2,\pi]$ in all cases except C4 where $[L_x,L_y,L_z]/h=[4\pi,2,2\pi]$, with $h$ being the channel half-width.
$M_x$, $M_z$ are the number of points for the equally-distanced grid in the streamwise and spanwise direction and $M_y$ are the number of points on a Chebyshev grid, including the 2 points at the walls.   
%Lengths in
%wall-units are indicated by  $[L_x^+,L_y^+,L_z^+]$.
$N_x$, $N_z$ are the number of ($\pm$) Fourier components after dealiasing and $N_y=M_y-2$.
%$R = h U_w / \nu$ is the bulk velocity Reynolds number according to the velocity at the wall $U_w$, the viscosity $\nu$ and the channel half-width $h$. 
All cases are run with $dt=0.008$ and $Re=3000$(C)/$Re=3250$(P), except for the C1 ensemble where $Re=1500$ and $dt=0.0125$.  The average execution time (on CPU or GPU) per time-step over a simulation with 3000 time-steps is denoted as $\bar{T}^{CPU/GPU}_{dt}$.}
%$R_{\tau} =  h u_{\tau} / \nu$ 
%is the Reynolds number of the simulation based on the friction velocity $u_{\tau} \sqrt{ \nu \left.\df U /\df y\right|_{\rm w}}$,where $\left.\df U/\df y\right|_{\rm w}$ is the shear at the wall
%.
%$\Delta x^+$ and $ \Delta z^+$ denote the average streamwise and spanwise grid spacing in wall units.}%(Add $Re_{bulk}$)}
%and $[L_x^+$,$L_z^+]$ is the channel size in wall units.}
%%\footnotesize\rm
\centering\vspace{.8em}
\begin{tabular}{@{}*{6}{c}}
\break
 Abbreviation  &$[M_x \times M_z \times M_y]$&$N_x\times N_z\times N_y$& $\bar{T}^{CPU}_{dt}$ & $\bar{T}^{GPU}_{dt}$ \\
 C1   & $[72\;,72,\;63]$&$47\times 47\times 61$&  $0.1620$ & $0.0606$  \\
 P1   & $[72\;,72,\;63]$&$47\times 47\times 61$& $0.1519$ & $0.0624$  \\
 C2   & $[108\;,108,\;73]$&$71\times 71\times 71$& $0.5129$ & $0.1308$ &\\
 P2   & $[108\;,108,\;73]$&$71\times 71\times 71$& $0.4992$ & $0.1366$ &\\
 C3  & $[144\;,144,\;83]$&$ 95 \times 95\times 81$& $1.1358$ & $0.2821$  \\
 P3  & $[144\;,144,\;83]$&$ 95 \times 95\times 81$& $1.1536$ & $0.2922$  \\
 C4  & $[156\;,156,\;83]$ & $ 103 \times 103 \times 81$ & $1.4349$ & $0.3390$ \\
\end{tabular}
%\label{table:geometry}
\end{table}
\end{center}

Solutions for the two cases were calculated using an i9-9900k CPU for 3000 time steps. The timing benchmarks between the two cases show that C2 completes a time-step about 2.5 times faster than VK3 on average. This already is a considerable improvement given the identical conditions of the test. The tradeoff paid for this speed boost is found in the respective memory utilization, where VK3 uses about ~500 MB, whereas C2 occupies about ~2500 MB. Additional cases are tested to document the dependence of $\bar{T}_{dt}$ as the number of grid points changes. For the DNSLab cases we observe an increase in $\bar{T}_{dt}$ that is steeper than linear (Fig.\ref{fig:Tdt}). On the other hand, cases C1-C4 run on GPU (Titan Z Black) exhibit a linear behaviour, suggesting a better scaling for our code across cases with sizes that fit in the GPU memory. %{\color{red} Memory requirements per simulation realization become less prohibitive with the ensemble solver.}

The costly part of the time integration in VK is the pressure correction. In the 3000 time-steps, their code spends approximately $80\%$ of the time on the correction and projection stages, with the majority of it owned to the bicgstab.m function solving the pressure Laplacian. For our code, the rkstep.m is the most time-consuming, particularly the multiplication of the precalculated matrices with the current $v$ and $\eta$ and the respective advection terms to obtain their new values.  

% the calculation of the new $v$ and $\eta$ %which require two matrix multiplications. 

\begin{center}
\begin{table}
%\caption{}
\caption{\label{table:grid1}Simulation parameters for VK cases.  
The channel  size in the streamwise, wall-normal and spanwise  directions is $[L_x,L_y,L_z]/h=[2\pi,2,\pi]$, where $h$ is  the half-width.
The number of points for the finite difference grid in the streamwise, spanwise and wall-normal directions are $N_x$, $N_z$ and $N_y$ (excluding the 2 points at the top and bottom wall).
%$M_x$, $M_z$ are the number of points for the finite difference grid in the streamwise and spanwise direction and $M_y+2$ are the number of points on the finite difference grid in the , including the 2 points at the walls.   
%Lengths in
%wall-units are indicated by  $[L_x^+,L_y^+,L_z^+]$.
%$N_x$, $N_z$ are the number of Fourier components after dealiasing and $N_y=M_y$ is the number of inner Chebyshev points.
%$R = h U_c / \nu$ is the bulk velocity Reynolds number according to the centerline velocity $U_c$, the viscosity $\nu$ and the channel half-width $h$. 
The average CPU execution time per time-step over a simulation with 3000 time-steps is denoted as $\bar{T}_{dt}$.}
%$R_{\tau} =  h u_{\tau} / \nu$ 
%is the Reynolds number of the simulation based on the friction velocity $u_{\tau} \sqrt{ \nu \left.\df U /\df y\right|_{\rm w}}$,where $\left.\df U/\df y\right|_{\rm w}$ is the shear at the wall
%.
%$\Delta x^+$ and $ \Delta z^+$ denote the average streamwise and spanwise grid spacing in wall units.}%(Add $Re_{bulk}$)}
%and $[L_x^+$,$L_z^+]$ is the channel size in wall units.}
%%\footnotesize\rm
\centering\vspace{.8em}
\begin{tabular}{@{}*{6}{c}}
\break
 Abbreviation  &$N_x\times N_z\times N_y$& $\bar{T}^{CPU}_{dt}$ \\
 VK1   &$48\times 48\times 48$& $0.2335$   \\
 VK2   &$60\times 60\times 60$& $0.5505$ \\
 VK3  &$ 72 \times 72\times 72$& $1.2808$   \\
 VK4  &$ 84 \times 84\times 84$& $2.5256$   \\
\end{tabular}
%\label{table:geometry}
\end{table}
\end{center}

The greatest boost accomplished when comparing the GPU with CPU execution on our code is seen in the fft2 calculations. There, total time spent to perform the ffts is reduced to the 1/6th of that required on the CPU. An improvement of $3.5\times$ is noted on the rkstep where the matrix multiplications responsible for inverting and updating the field to the next intermediate step are executed. Finally, a nearly $2.5\times$ reduction is seen in the evaluation of the nonlinear advection term, where matrix and point by point multiplications occur. These last operations are also more computationally intensive since here the full, rather than the dealiased, degrees of freedom come into play.
To investigate a possible reduction of execution time in the VK code versions of the bicgstab function should be ported efficiently to GPU, however we are not aware of any such version available for MATLAB at the time, while the DNSLABIB uses different methods to obtain the pressure.

\section{Growth functional and linearized adjoint equations of streamwise-mean flow dynamics}

The problem of optimal perturbations is solved using the forward - adjoint formulation of \citep{Luchini-2000} for disturbances of a specified $k_x$. We define the energy growth as the ratio of initial to target time energy density, 
\begin{equation}
G(T_{opt}) = E(T_{opt})/E(0),
\end{equation}
for perturbations of initial kinetic energy density  $E=1/(2V) \int_V \u ' \cdot \u' = 1$ in the total domain volume V. This functional is augmented with the Navier-Stokes equations linearized about a streamwise-mean $\U = (U,V,W)$,
\begin{align} \partial_t \u'  =  - [ (\U \cdot \nabla) \u ' +  (\u' \cdot \nabla) \U  + \nabla p' - \frac{1}{Re} \Delta \u'] & \\
\nabla \cdot \u' = 0 & , \label{eq:lin_NS}
\end{align}
using Lagrange multipliers $\lambda = (\lambda_1,\lambda_2,\lambda_3), b$,
\begin{align}
\tilde{G}(T_{opt}) = & -\int_0^{T_{opt}} \int_V \lambda^T  [\partial_t \u' + (\U \cdot \nabla) \u ' +  (\u' \cdot \nabla) \U  + \nabla p' - \frac{1}{Re} \Delta \u'] \nonumber \\ & - \int_0^{T_{opt}} \int_V b \nabla \cdot \u' + G(T_{opt})
\end{align}
We derive the adjoint equations
\begin{align} \partial_t \mathbf{\lambda} =   [ (\U \cdot \nabla) \mathbf{ \lambda } +   (U_i \nabla \lambda_i)  + \nabla b + \frac{1}{Re} \Delta \mathbf{\lambda}]& \\
\nabla \cdot \lambda=0& . \label{eq:lin_NS}
\end{align}
 by varying the functional with respect to the forward equations and enforcing the conditions that render it stationary. 
 %Under this assumption 
 
For optimal perturbations the initial condition of the adjoint is the final state of the forward integration, which is then evolved backwards over the same time interval. The new forward initial condition is the normalized adjoint final state. The modes of this coupled dynamics are normal, therefore we can obtain higher order optimals by a Gram-Schmidt orthogonalization procedure. 

% FFT bottleneck
% hindered by memory deallocations
% sparse matrix multiplications remain efficient even at the largest ensemble sizes.

%% References


\bibliographystyle{elsarticle-num}
\bibliography{../../bibfile/basic_references}






% \end{thebibliography}

\end{document}